\documentclass[traditabstract]{aa}

\pdfoutput=1

\usepackage{natbib}
\usepackage{subcaption}
\usepackage{amssymb,amsfonts}
\usepackage{amsmath}
\usepackage{mathtools}
\usepackage{url}
\usepackage{hyperref}
\usepackage{txfonts}
\usepackage{graphics}

\newcommand{\bfOmega}{\mathbf{\Omega}}
\newcommand{\tkq}{\mathcal{T}^K_Q}
\newcommand{\jkq}{J^K_Q}
\newcommand{\ylm}{Y_{\ell m}}
\newcommand{\ylmprime}{Y_{\ell^{\prime} m^{\prime}}}
\newcommand{\Nomega}{N}

\begin{document}

\title{Improved near optimal angular quadratures for polarised
  radiative transfer in 3D MHD models\thanks{The tables mentioned in
    Sect. \ref{sec:concl} can be found in electronic form at the CDS
    via the anonymous ftp to \url{cdsarc.u-strasbg.fr} (130.79.128.5)
    or via \url{http://cdsarc.u-strasbg.fr/viz-bin/cat/J/A+A/645/A101}}}

\author{Jaume Jaume Bestard\inst{1,2} \and Ji\v{r}\'{\i}
  \v{S}t\v{e}p\'an\inst{3} \and Javier Trujillo Bueno\inst{1,2,4}}

\institute{ Instituto de Astrof\'{\i}sica de Canarias, V\'{\i}a
  L\'actea s/n, E-38205 La Laguna, Tenerife,
  Spain. \email{jjaumeb@iac.es} \and Departamento de Astrof\'{\i}sica,
  Universidad de La Laguna (ULL), E-38206 La Laguna, Tenerife, Spain
  \and Astronomical Institute ASCR, v.v.i., Ond\v{r}ejov, Czech
  Republic. \email{jiri.stepan@asu.cas.cz} \and Consejo Superior de
  Investigaciones Cient\'{\i}ficas, Spain. \email{jtb@iac.es} }

\date{Received 14 September 2020 / Accepted 22 November 2020}

\abstract{ Accurate angular quadratures are crucial for the numerical
  solution of three-dimensional (3D) radiative transfer problems,
  especially when the spectral line polarisation produced by the
  scattering of anisotropic radiation is included. There are two
  requirements for obtaining an optimal quadrature and they are
  difficult to satisfy simultaneously: high accuracy and short
  computing time. By imposing certain symmetries, we were recently
  able to derive a set of near optimal angular quadratures. Here, we
  extend our previous investigation by considering other
  symmetries. Moreover, we test the performance of our new quadratures
  by numerically solving a radiative transfer problem of resonance
  line polarisation in a 3D model of the solar atmosphere resulting
  from a magneto-hydrodynamical simulation. The new angular
  quadratures derived here outperform the previous ones in terms of
  the number of rays needed to achieve any given accuracy.  }


\titlerunning{Improved near optimal angular quadratures}
\authorrunning{J. Jaume Bestard et al.}

\maketitle


\section{Introduction\label{sec:intro}}

High-precision numerical integration is essential for obtaining
reliable results when solving radiative transfer (RT) problems. An
inaccurate angular integration of the radiation field can lead to
serious errors that affect the solution of the statistical equilibrium
equations (SEE), especially for problems involving the polarisation of
spectral lines due to the scattering of anisotropic radiation.

In a previous paper \citep[][hereafter, Paper~I]{2020A&A...636A..24S},
we formulated the problem of optimal angular quadratures for polarised
radiative transfer. Under certain considerations of symmetry, we found
a new set of quadratures that perform better than the most commonly
used ones, meaning that they provide the same analytical accuracy by
using a smaller number of quadrature rays. In the present paper, we
generalise the symmetry constraints with the aim of finding even
better angular quadratures. Moreover, we perform comparative tests of
the quadrature sets using three-dimensional (3D)
radiation-magnetohydrodynamical (r-MHD) models of the solar
atmosphere.

In Sect.~\ref{sec:deriv}, we briefly review our definition of an
optimal quadrature and the most relevant equations. Then we devise
several groups of quadrature symmetries. Finally, we discuss an
efficient numerical algorithm for a practical construction of the
quadratures. In Sect.~\ref{sec:results}, we calculate a new set of
quadratures with different symmetries and we study their accuracies,
both analytically and via a 3D out of local thermodynamic equilibrium
(non-LTE) radiative transfer calculation in a realistic r-MHD
model. Finally, in Sect.~\ref{sec:concl}, we discuss the results and
present our conclusions.


\section{Derivation of the near optimal quadratures\label{sec:deriv}}


\subsection{Formulation of the problem\label{ssec:formul}}

This section briefly recalls the problem that we addressed in Paper~I
(where the reader can find all the necessary definitions and
derivations). Below, we only provide a summary of the key steps in the
development of optimal angular quadratures.

The angular integration of the Stokes parameters
$(I,Q,U,V)=(I_0,I_1,I_2,I_3)$ \cite[e.g.][]{2004ASSL..307.....L} over
the unit sphere
${\mathbb{S}^2=\{ \bfOmega \in \mathbb{R}^3 : ||\bfOmega||=1 \}}$ is
usually done using an angular quadrature
$\{w_i,\vec\Omega_i\}_{i=1}^N$, which is defined as a set of $N$
positive weights, $w_i$, and directions, $\vec\Omega_i$. Here, we
consider the common definition of the spherical coordinates,
$\bfOmega=(\theta,\phi),$ in which the inclination, $\theta$, is
measured from the positive $z$ axis and the azimuth, $\phi$, with
respect the positive $x$ axis.  We showed in Paper~I that expanding
the Stokes parameters into a truncated basis of the real-valued
spherical harmonics (SH),
\begin{equation}
  \label{eq:expansion}
  I_j(\bfOmega) = \sum^{L}_{\ell=0}\sum^{\ell}_{m=-\ell} \left( I_j\right)_{\ell m} \ylm(\bfOmega) \,,
\end{equation}
together with the requirement that the quadrature allows an exact
calculation of the radiation field tensors, $J^K_Q$, for the Stokes
parameters expanded up to order $L$ (see Eq.~4 of Paper~I) leads to
the following set of non-linear equations:
\begin{equation}
  \begin{split}
  \label{eq:error_jkq}
  &  \sum^3_{j=0} \sum^{L}_{\ell=0}\sum^{\ell}_{m=-\ell} \left( I_j\right)_{\ell m} \int_{\mathbb{S}^2} \ylm(\bfOmega) \, \tkq(j,\bfOmega) \, d\bfOmega  \\
 =& \sum^3_{j^{\prime}=0} \sum^{L}_{\ell^{\prime}=0}\sum^{\ell^{\prime}}_{m^{\prime}=-\ell^{\prime}} \left( I_{j^{\prime}}\right)_{\ell^{\prime} m^{\prime}} \sum_{i=1}^{\Nomega} w_i \ylmprime(\bfOmega_i)\tkq(j^{\prime},\bfOmega_i) \, ,
  \end{split}
\end{equation}
which needs to be satisfied for any radiation field expansion
coefficients $\left( I_j\right)_{\ell m}$ and
$\left( I_{j^{\prime}}\right)_{\ell^{\prime} m^{\prime}}$. It follows
from this requirement that the quadrature needs to satisfy a simple
set of equations,
\begin{equation}
  \label{eq:system}
  y(j,\ell,m,K,Q) -  \sum_{i=1}^{\Nomega} w_i \ylm(\bfOmega_i)\tkq(j,\bfOmega_i) = 0 \, ,
\end{equation}
where the first term is a complex-valued integral,
\begin{equation}
  \label{eq:ylm}
  y(j,\ell,m,K,Q) = \int_{\mathbb{S}^2} \ylm(\bfOmega) \, \tkq(j,\bfOmega) \, d\bfOmega \, ,  
\end{equation}
which can easily be evaluated analytically. The set of Eqs.
(\ref{eq:system}), with all the possible combinations of
$(j,\ell,m,K,Q)$ up to $\ell\le L,$ form the system of equations for
the unknowns $w_i$ and $\bfOmega_i$. This system is non-linear and
non-convex and needs to be solved by numerical methods. The quadrature
with the lowest number of rays that can solve such a system of
equations is said to be optimal. Since we cannot provide an analytical
proof that any of our numerically discovered quadratures are optimal,
we refer to near optimal quadratures for the purposes of this work.

The number of equations in the system of Eqs.~(\ref{eq:system}) grows
proportionally to $(L+1)^2$. Since it is non-linear and non-convex and
its size rapidly grows with the required accuracy, the numerical
solution becomes increasingly more difficult as $L$ increases. It was
possible to find near optimal quadratures up to $L=15$, what is
sufficient for solar-like atmospheres using conventional numerical
techniques (see Sect.~3 of Paper~I). However, it would be more
difficult for larger $L$ values, as well as in the case of quadratures
with relaxed symmetries where the number of unknowns is significantly
larger. Therefore, we had to use a different strategy (see
Sect.~\ref{ssec:nea}) that is extremely efficient and without
convergence problems.

From a numerical perspective, it is worth mentioning that angular
symmetries of the quadratures can lead to an automatic fulfillment of
some of the equations in the system of equations (\ref{eq:system})
because of the analytical form of SH and $\mathcal{T}^K_Q$. Therefore,
such equations can be ignored in the numerical solution. This fact
further decreases the computing demands of the quadrature
calculation. For example, the equation defined by
$(j,\ell,m,K,Q)=(0,0,0,2,2)$ is fulfilled for any quadrature that is
rotationally symmetric with respect to the $z$ axis because of the
exponential factor in the azimuthal integral of the geometrical
tensor.

The problem of finding an optimal angular quadrature becomes greatly
simplified when we only account for the specific intensity, $I_0$. In
this case, the system of equations (\ref{eq:system}) that needs to be
solved are simply the spherical harmonics up to order $L$.


\subsection{Symmetries of the quadratures\label{ssec:symm}}

Since the SH basis functions have certain rotationally invariant
properties, it is also natural to impose several symmetries in the
derivation of the quadratures. Furthermore, as discussed in the
previous section, symmetries can considerably reduce the number of
equations that have to be solved as well as the number of unknowns. On
the other hand, imposing an ad hoc symmetry of the quadrature is an
assumption that may keep us from finding the more general and more
optimal quadratures. In this section, we study several symmetry
properties of the quadratures and we also consider the most general
and, therefore, the most numerically demanding case in which all the
symmetries are relaxed.

First of all, we briefly introduce the analytical description of the
group of symmetries considered in this work. We define a finite group
of rotations and reflections, namely, $\mathcal{G}\in O(3)$. A
quadrature with ray directions and weights,
$\{w_i,\bfOmega_i\}_{i=1}^{\Nomega}$, is said to be invariant under
$\mathcal{G}$ if, for any rotation or reflection $g\in \mathcal{G}$,
the quadrature meets the condition
\begin{equation}
  \label{eq:inv_cond}
  \sum^{\Nomega}_{i=1} w_i f(g^{-1}\bfOmega_i) = \sum^{\Nomega}_{i=1}w_if(\bfOmega_i) \, ,
\end{equation}
where $f(\cdot)$ represents a value of the integrand at a given ray
direction \citep{doi:10.1098/rspa.2009.0104}. The above condition
allows us to reduce the number of unknowns of the problem. In other
words, we can take advantage of the symmetries,
$\{g_k\}\in \mathcal{G,}$ to define the quadrature from a set of $N_g$
(with $N_g\le N$) generating nodes,
$\{\tilde{w}_i,\tilde{\bfOmega}_i\}_{i=1}^{N_g}$.  The rest of the
nodes over the unit sphere are found using the so-called orbits of the
quadrature \citep[see][for more details]{XIAO2010663}. By orbit
$\mathcal{O}_{i},$ associated with the generating node $i$, we mean a
set of nodes that can be obtained by applying any set of the group
transformations $\{g_k\},$ and having the same weight, $w_i$, as the
generating node.

The quadratures presented in Paper~I were derived with the assumption
that the nodes were invariant under $90^{\circ}$~rotations around the
$z$ axis and reflections with respect to the $xy$ plane. Using the
group notation, this can be formally defined as follows:
\begin{equation}
  \label{eq:s1}
  \begin{array}{l}
    S_1: \left\{g_z(90^{\circ}),\, g_{xy}\right\},\\\\
    \left\{\tilde{w}_i,~ 0<(\tilde{\theta}_i,\tilde{\phi}_i)<\dfrac{\pi}{2}\right\}^{N_g}_{i=1}.
  \end{array}
\end{equation}
The generating nodes are defined in the $xyz$-positive octant and the
orbit to generate the other nodes uses the group of symmetries
$S_1$. In this case, the number of generating nodes $N_g$ is the
number of nodes per octant, ${N_g=n=\Nomega/8}$ (i.e. we have $3n$
unknowns in total).

In this paper, we extend the previous work to quadratures with other
angular symmetries. Firstly, the symmetry group labelled as $S_2$ can
be defined as:
\begin{equation}
  \label{eq:s2}
  \begin{array}{l}
    S_2: \left\{g_z(90^{\circ}),\, g_{xy},\, g_{xz},\, g_{yz}\right\},\\\\ 

    \left\{\tilde{w}_i,~ 0<\tilde{\theta}_i<\dfrac{\pi}{2},~ 0<\tilde{\phi}_i\le\dfrac{\pi}{4}\right\}^{N_g}_{i=1}.
  \end{array}
\end{equation}
This is a group with rotational symmetry around the $z$ axis and
reflection symmetry with respect to all the coordinate planes. We
point out that the well-known tensor product quadrature formed by the
Gaussian quadrature in the cosine of inclinations and the equally
spaced trapezoidal quadrature in the azimuths (hereafter, the GT
quadrature) is a special case of the quadratures with $S_2$
symmetries. The generating nodes of the $S_2$ quadratures are defined
in a half of the positive octant.

We also define the symmetry group $S_3$ as:
\begin{equation}
  \label{eq:s3}
  \begin{array}{l}
    S_3: \left\{g_{x}(90^{\circ}),\, g_{y}(90^{\circ}),\, g_{z}(90^{\circ}),\, g_{xy},\, g_{xz},\, g_{yz}\right\}, \\\\
    \small\left\{\tilde{w}_i,~ 0<\tilde{\theta}_i\le \arcsin{\frac{1}{\sqrt{2-\sin^2{\tilde{\phi}_i}}}},~ 0<\tilde{\phi}_i\le\dfrac{\pi}{4}\right\}^{N_g}_{i=1} ~,
  \end{array}
\end{equation}
which has, in addition, the rotational symmetry with respect to the
$x$ and $y$ axes. In the case of the $S_3$ quadratures, the generating
nodes are enclosed in a sextant of the positive octant. This is the
region delimited by $\hat{x}=\hat{y}$ and $\hat{x}=\hat{z,}$ where
$\hat{x}$, $\hat{y}$ and $\hat{z}$ are the direction cosines of each
axis. The upper limit for $\tilde\theta_i$ can be obtained from simple
trigonometry using the fact that ${\hat{x}^2+\hat{y}^2+\hat{z}^2=1}$.

Aside from the above types of quadratures, we also define
non-symmetric quadratures, $S_0$, which do not generally have any
rotational or reflection symmetry. These quadratures have a large
amount of unknowns, namely $3\Nomega$, and none of the
Eqs.~(\ref{eq:system}) are automatically satisfied. This kind of
quadrature is the most numerically challenging and more advanced
techniques need to be implemented to find them.


\subsection{Node elimination algorithm\label{ssec:nea}}

The method used to calculate the $S_0$, $S_2$, and $S_3$ quadratures
differs from the one used to calculate the $S_1$ quadratures in
Paper~I. Here, we implement the so-called node elimination algorithm
proposed by \cite{XIAO2010663}. This algorithm combines Newton's
method for finding the roots of Eqs.~(\ref{eq:system}) and a node
elimination scheme. The main idea behind this algorithm is to
initialise the minimisation problem with an approximate solution and
then successively remove particular rays while keeping the exactness
of the quadrature up to order $L$ using Newton's method. In this
process, we eventually arrive at a quadrature with an optimal number
of rays.  This node elimination scheme helps to solve the major
problem of the numerical solution, which is the sensitivity to the
initial guess and the tendency to end up in a sub-optimal
solution. Our numerical experiments suggest that if the initial
quadrature has sufficient rays, the algorithm becomes very robust and
solutions with different initialisations converge to quadratures with
the same near optimal or perhaps optimal number of rays.  We briefly
summarise below the main steps of our implementation of the algorithm
in order to find the generating rays of the quadrature of order $L$.

Firstly, we initialise the quadrature using a large number of nodes,
$N\sim (L+1)^2$. The rays and weights can be those of the GT
quadrature or randomly distributed over the unit sphere. In the case
of a random initialisation, we use Newton's method to find the exact
quadrature. We then sort the generating nodes in increasing order of
their significance\footnote{See \citet{XIAO2010663} for the definition
  of ray significance.}, $s$, and remove the least significant node
from the quadrature. We use Newton's method to find an exact
quadrature. If an exact quadrature is found, we keep the process of
sorting by significance and removing node by node.  Otherwise, we
reestablish the removed node and remove the next one with higher $s$,
if there is any, and we apply again Newton's method. If there are no
more nodes to eliminate, we conclude that we have thus found a near
optimal quadrature.

Given that the system of equations (\ref{eq:system}) is solved
numerically, the condition of exactness is the same as in Paper~I,
namely:
\begin{equation}
  \label{eq:condition}
  \underset{j,\ell,m,K,Q}{\max}
  \small{\left\|y(j,\ell,m,K,Q) - \sum^{\Nomega}_{i=1} w_i \ylm(\bfOmega_i) \tkq(j,\bfOmega_i)\right\| < 10^{-15} \,},
\end{equation}
for any combination of the $(j,\ell,m,K,Q)$ indices.


\section{Results\label{sec:results}}

\subsection{Analytical accuracy}

\begin{figure}
  \begin{center}
    \includegraphics[width=1\columnwidth]{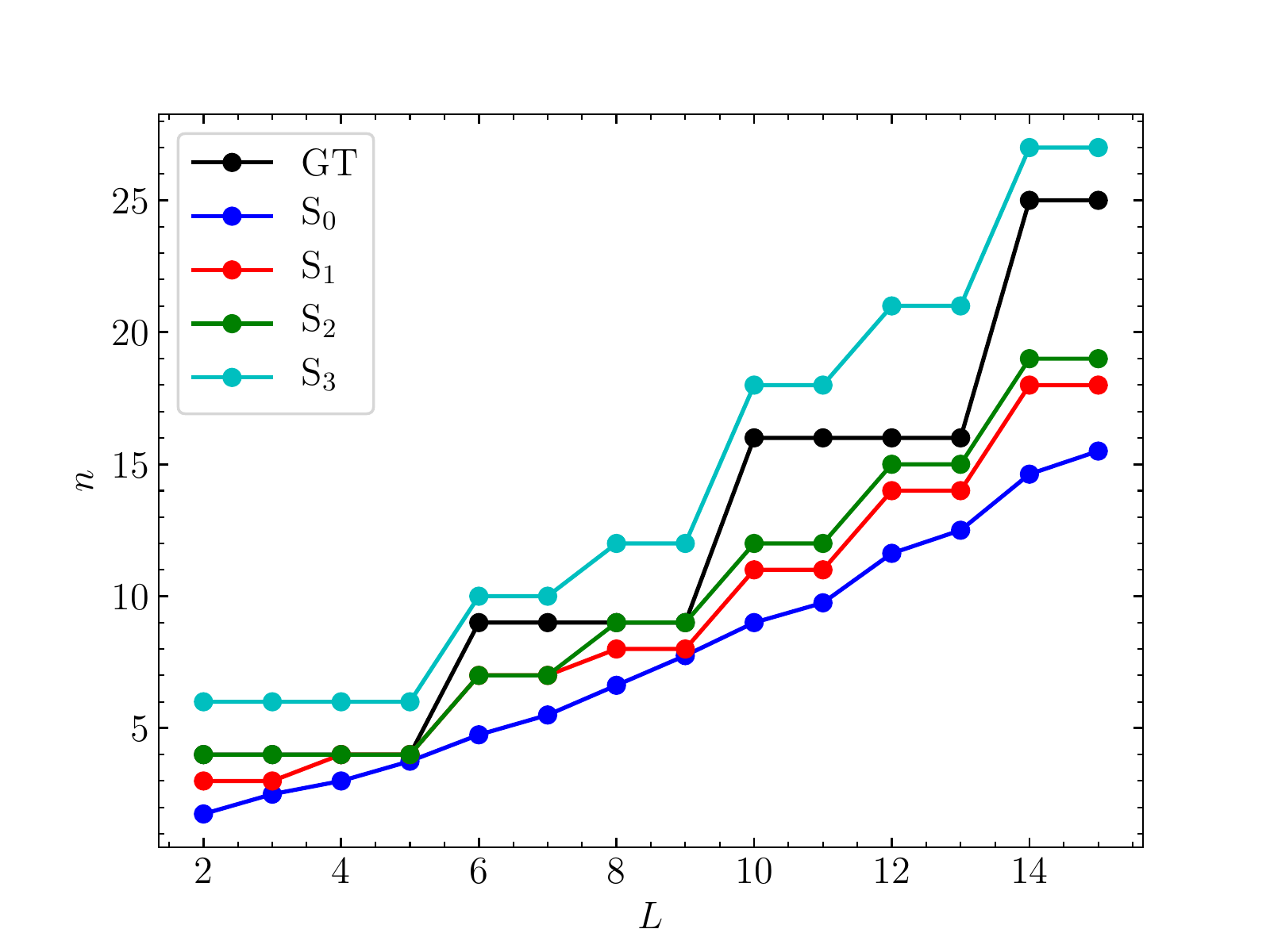}
  \end{center}
  \caption{Number of rays per octant, $n=N/8$, as a function of the
    order $L$ of the quadrature. We note that, in the case of the
    $S_0$ quadratures, $n$ is generally not an integer. See the text
    for more details.}
  \label{fig:nrays}
\end{figure}

\begin{figure*}[]
  \begin{subfigure}[b]{0.32\textwidth}
    \includegraphics[width=\columnwidth]{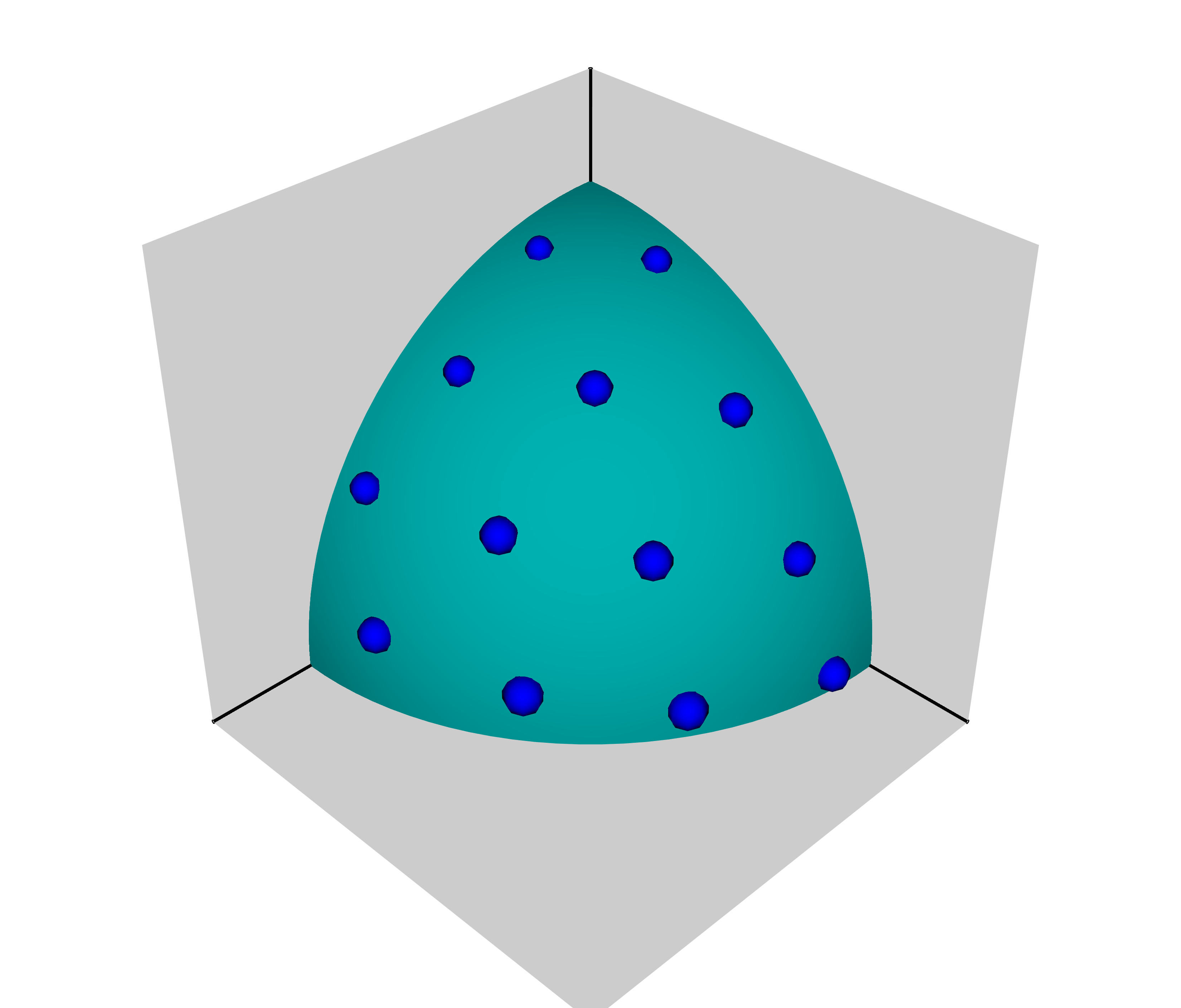}
    \caption{$S_0$}
  \end{subfigure}
  \begin{subfigure}[b]{0.32\textwidth}
    \includegraphics[width=\columnwidth]{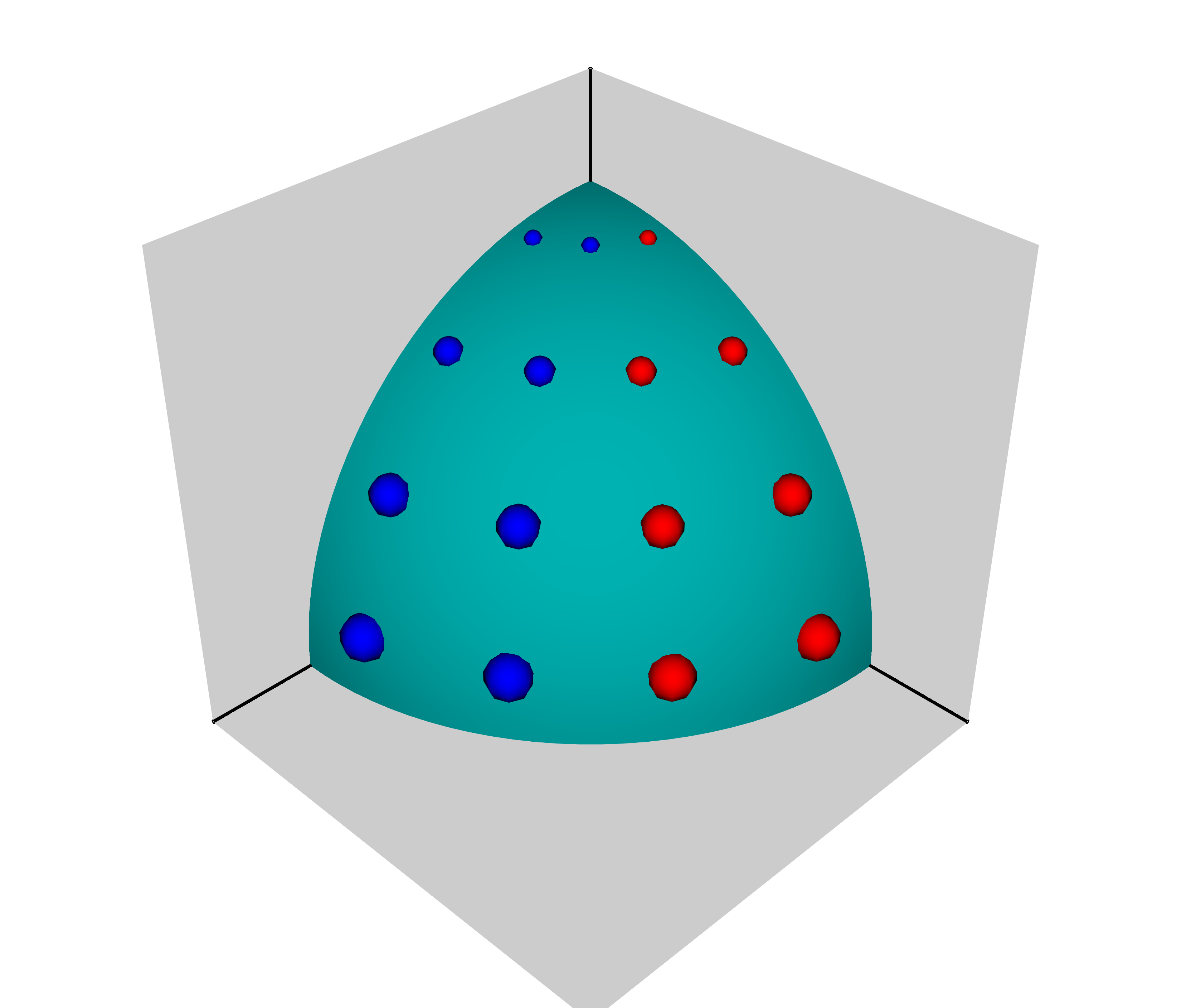}
    \caption{$S_2$}
  \end{subfigure}
  \begin{subfigure}[b]{0.32\textwidth}
    \includegraphics[width=\columnwidth]{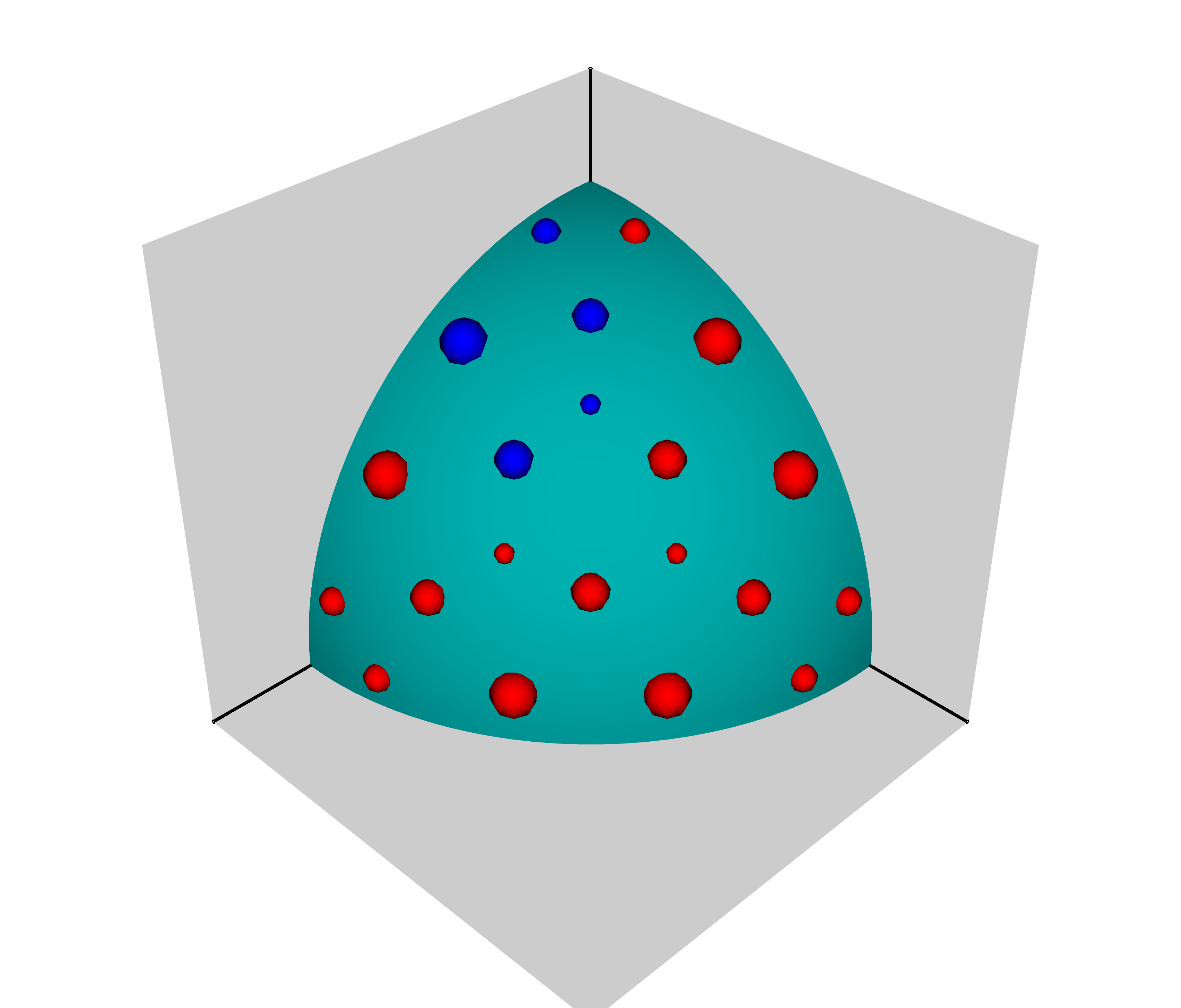}
    \caption{$S_3$}
  \end{subfigure}
  \caption{Examples of near optimal angular quadratures of order
    $L=13$. Left panel: $S_0$ quadrature with a total of 100
    independent rays. Central panel: $S_2$ quadrature with 8
    independent rays and 120 rays in total. Right panel: $S_3$
    quadrature with 5 independent rays and 168 rays in total. We note
    that although only one octant is shown, the independent rays in
    the left panel are distributed over the whole sphere. The blue
    points indicate the generating nodes while the red points are on
    their orbits.}
  \label{fig:oaq}
\end{figure*}

Using the node elimination algorithm, we have found near optimal
quadratures of the types $S_0$, $S_2,$ and $S_3$ for an increasing
order of accuracy, $L$. As in Paper~I, we show in Fig.~\ref{fig:nrays}
the number of rays per octant of the different quadratures for the
polarised case.

The main result shown in this figure is that the $S_0$ quadrature
type, that is, the most general one that lacks any rotational or
reflection symmetry, is more efficient than the quadratures found in
Paper~I (the $S_1$ type, see the blue line in
Fig.~\ref{fig:nrays}). For ${L=15}$, the $S_1$ quadrature needs about
16\,\% more rays than the $S_0$ quadrature.  We also applied the node
elimination algorithm to the $S_1$ quadratures studied in Paper~I
(shown with a red line in Fig.~\ref{fig:nrays}) and we verified that
the number of rays is identical as previously reported.

The $S_2$ quadratures are typically slightly worse in terms of number
of rays than $S_1$ but the difference is quite small up to $L=15$. The
GT quadratures also belong to the $S_2$ family as a special case, but
their analytical accuracy is significantly worse (see the green and
black lines in Fig.~\ref{fig:nrays}).

The worst performance is found in the case of the $S_3$ quadratures
(see the cyan line in Fig.~\ref{fig:nrays}). These quadratures have
the highest level of symmetries and, therefore, they are easy to find
numerically. However, given that their accuracy is even worse than the
GT quadratures, which can be generated without any numerical effort,
there is no practical benefit in using the $S_3$ quadratures, apart
from the fact that the rays are more equally distributed over the unit
sphere than in the case of the GT quadratures.

Figure~\ref{fig:oaq} shows an example of three near optimal
quadratures $S_0$, $S_2,$ and $S_3$ of order $L=13$. We note that, in
the case of the $S_0$ quadrature (left panel), the rays in the first
octant do not determine the rays in the other octants due to the lack
of symmetries. The $S_0$ quadratures need fewer rays than the other
quadratures to achieve a given accuracy. This is because all the rays
over the unit sphere are independent and can be more efficiently
distributed. For the other quadratures, the independent (generating)
rays are limited to a small region and used redundantly.

\subsection{Tests in a 3D non-LTE simulation}

\begin{figure*}
  \begin{subfigure}[b]{0.33\textwidth}
    \includegraphics[width=\textwidth]{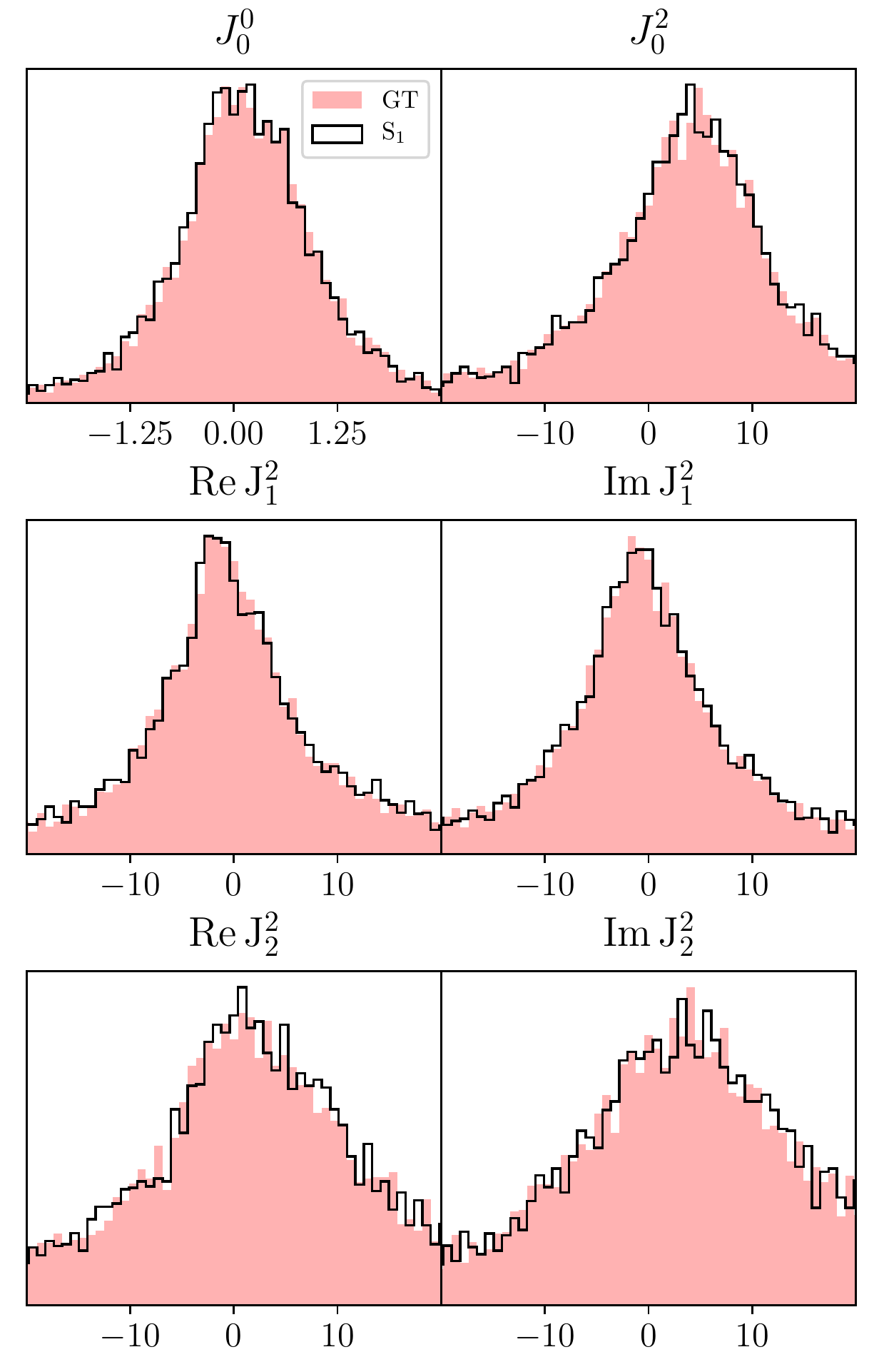}
    \caption{$L=9$}
  \end{subfigure}
  \begin{subfigure}[b]{0.33\textwidth}
    \includegraphics[width=\textwidth]{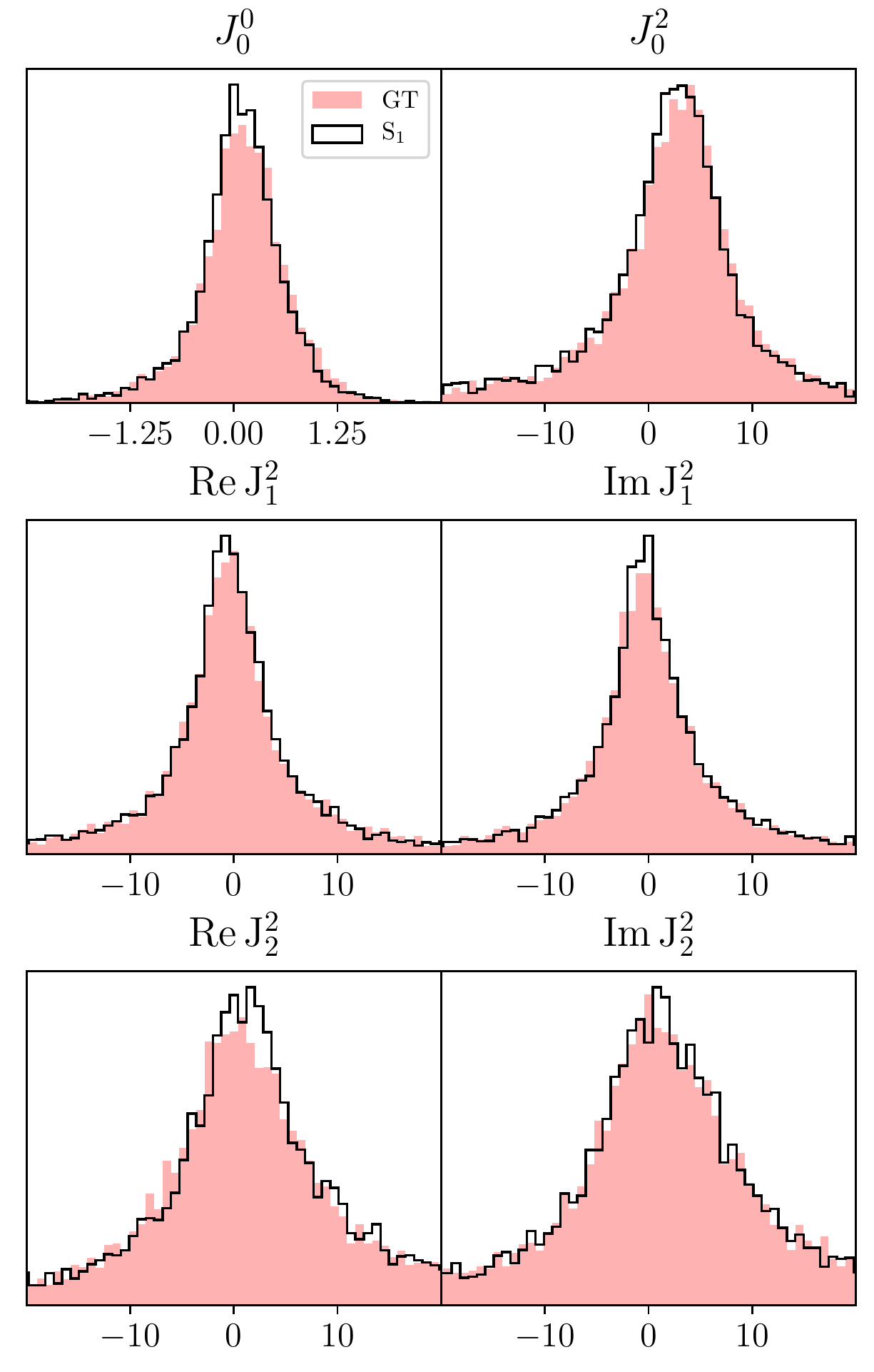}
    \caption{$L=13$}
  \end{subfigure}
  \begin{subfigure}[b]{0.33\textwidth}
    \includegraphics[width=\textwidth]{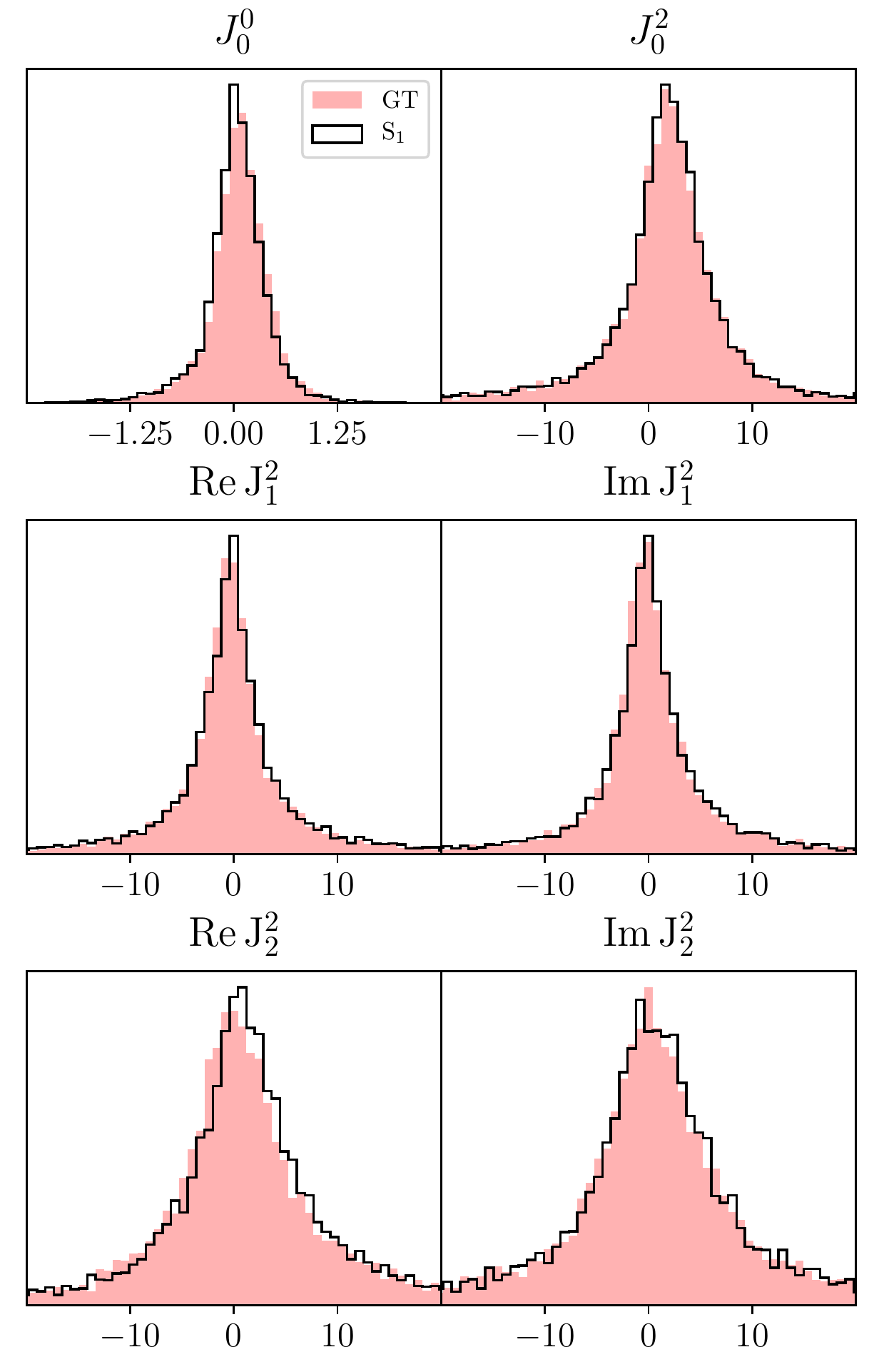}
    \caption{$L=15$}
  \end{subfigure}
  \caption{Histograms showing the relative errors in the percentage of
    the radiation field tensors,
    $(J^K_Q-\left[J^K_Q\right]_{\mathrm{ref}})/\left[J^K_Q\right]_{\mathrm{ref}}$,
    with respect to the reference solution,
    $\left[J^K_Q\right]_{\mathrm{ref}}$, at the corrugated surface
    where the optical depth at the line centre is unity along the disc
    centre line of sight. The different panels show: (a) GT $3\times3$
    and S1 with 8 rays per octant, (b) GT $4\times4$ and S1 with 14
    rays per octant, and (c) GT $5\times5$ and S1 with 18 rays per
    octant.}
  \label{fig:histograms_s1}
\end{figure*}

\begin{figure*}
  \begin{subfigure}[b]{0.33\textwidth}
    \includegraphics[width=\textwidth]{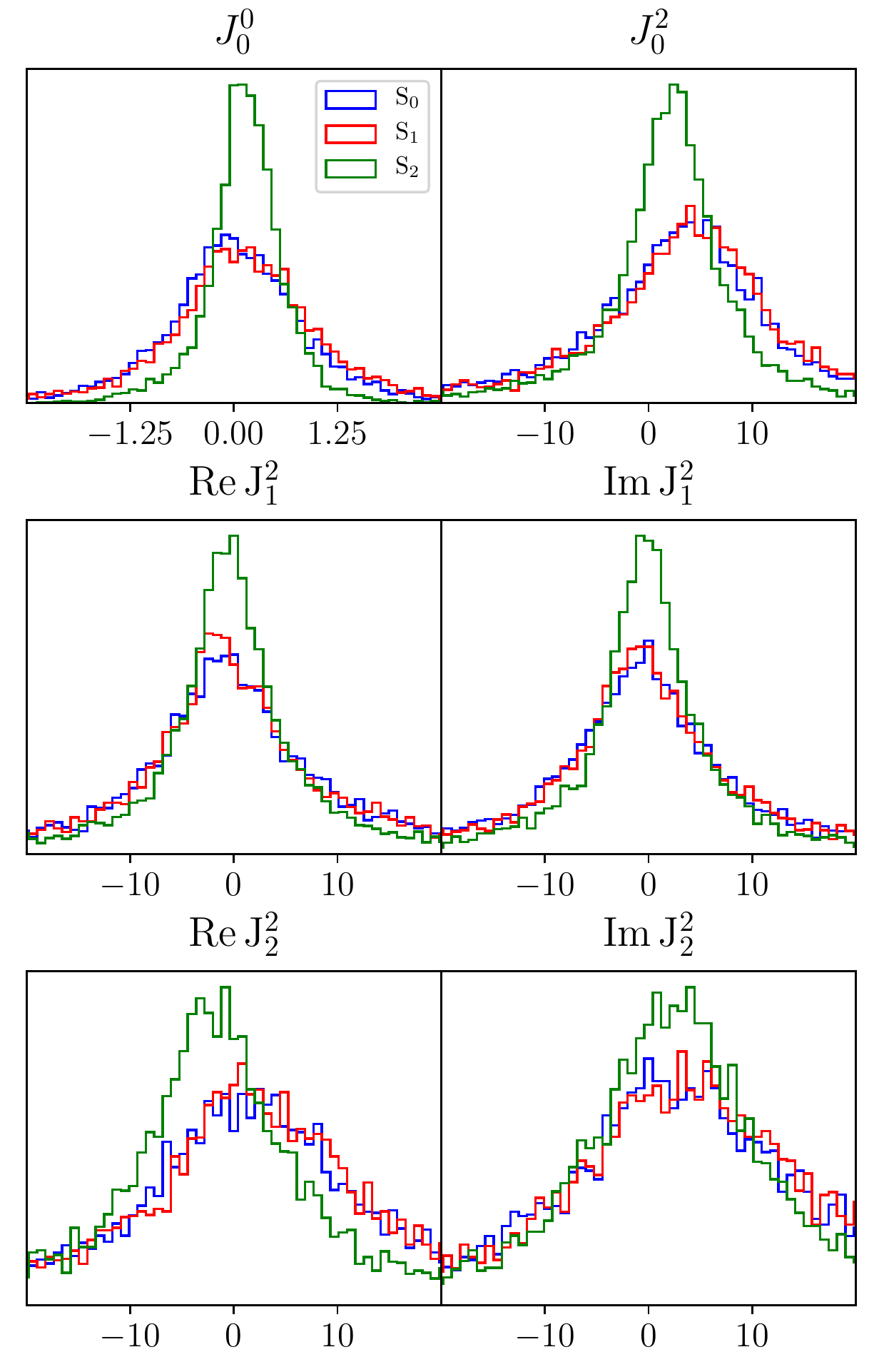}
    \caption{$L=9$}
  \end{subfigure}
  \begin{subfigure}[b]{0.33\textwidth}
    \includegraphics[width=\textwidth]{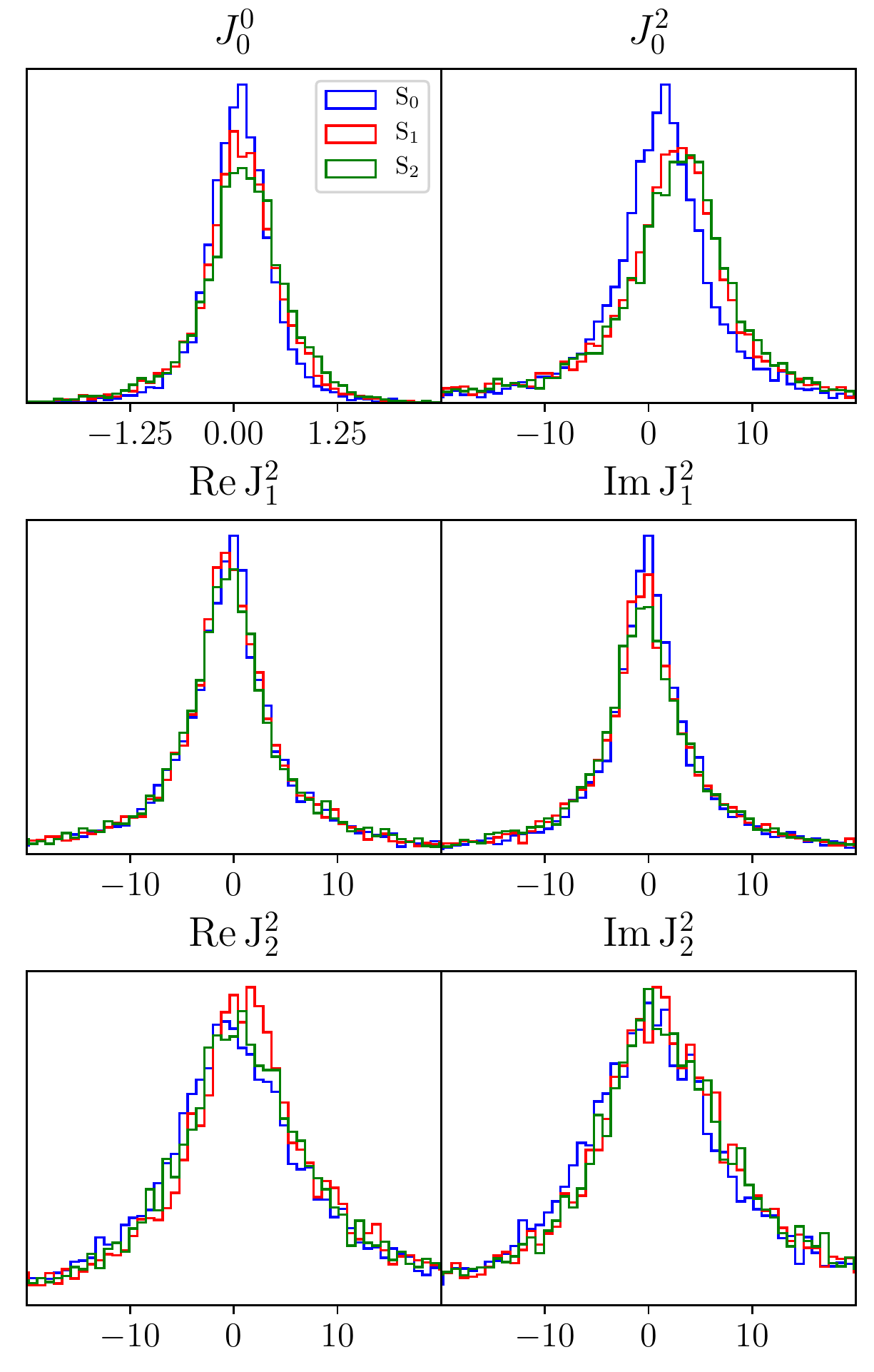}
    \caption{$L=13$}
  \end{subfigure}
  \begin{subfigure}[b]{0.33\textwidth}
    \includegraphics[width=\textwidth]{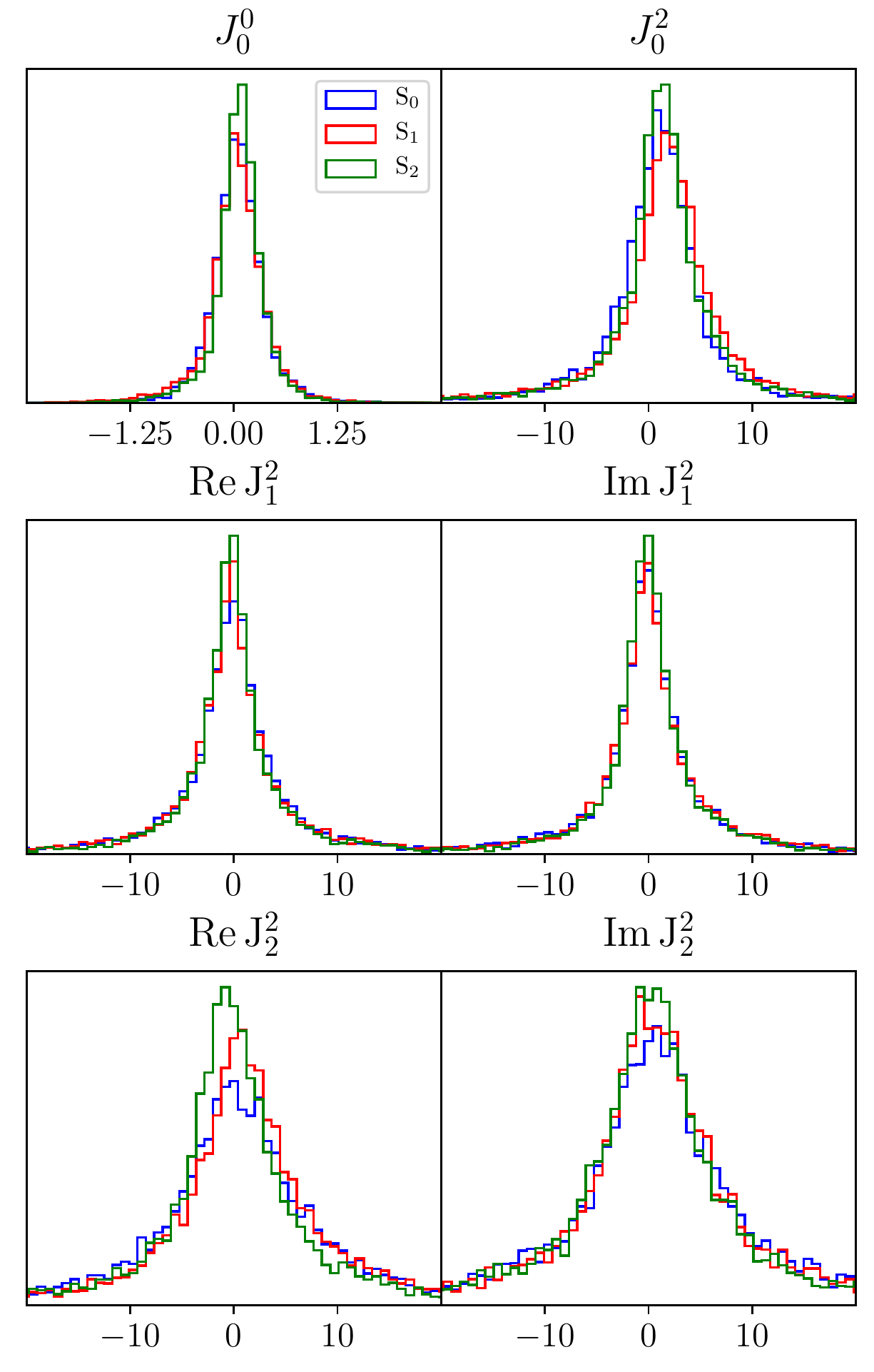}
    \caption{$L=15$}
  \end{subfigure}
  \caption{Histograms showing the relative errors in percentage of the
    radiation field tensors,
    $(J^K_Q-\left[J^K_Q\right]_{\mathrm{ref}})/\left[J^K_Q\right]_{\mathrm{ref}}$,
    at the height where the optical depth at the line centre is unity
    along the disc centre line of sight. The different panels show the
    histograms for different truncation orders of the spherical
    harmonics expansion for the $S_0$, $S_1,$ and $S_2$ quadratures
    with the following number of rays in total, respectively: (a) 62,
    64 and 72 rays, (b) 100, 112 and 120 rays, and (c) 124, 144 and
    152 rays in total.}
  \label{fig:histograms_s0_s1_s2}
\end{figure*}

In the derivation of the near optimal quadratures, we
restrict the expansion of the Stokes parameters up to a certain order
$L$ of the SH. In fact, this always leads to approximate
quadratures. Indeed, the smoother the angular variation of the
radiation field, the more accurate the expansion of
Eq.~(\ref{eq:expansion}).

In Paper~I, we presented an example of how the amplitudes of the
expansion coefficients $(I_j)_{\ell}$ decrease with the order $\ell$
(see Eq.~16 and Fig.~2 therein). This can be used for obtaining an
empirical determination of the maximum order $L$ required in a
particular type of model.

There is no a priori guarantee that two different types of quadratures
with the same order $L$ will lead to the same accuracy in 3D non-LTE
calculations in which Eq.~(\ref{eq:expansion}) provides merely an
approximation of the radiation field. Some of the quadratures may
provide better accuracy than others when integrating the radiation
field with a non-negligible contribution of the $\ell>L$ modes. This
is a model-dependent problem and we study it in this section using
non-LTE radiative transfer in a 3D model of the solar atmosphere.

In order to check the actual accuracy on the $J^K_Q$ tensors using the
new angular quadratures, we solved the non-LTE radiative transfer
problem for polarised light in a 3D model resulting from a MHD
simulation \citep{2011A&A...531A.154G,2016A&A...585A...4C}, using the
radiative transfer code PORTA\footnote{See the public version of PORTA
  at \url{https://gitlab.com/polmag/PORTA}}
\citep{2013A&A...557A.143S}. Due to the significant computing demands
of 3D non-LTE radiative transfer and the high number of different
quadratures to test, we decided to reduce the horizontal spatial
resolution of the model grid by a factor of $6$ at the expense of
having more abrupt spatial variations on the physical parameters. The
horizontal spatial resolution is approximately $240$~km after the
degradation. The vertical resolution remains the same as in the
original model. The spectral line considered here is the strong
resonance transition of neutral calcium at $422.7\,\mathrm{nm}$. This
transition can be modelled using a two-level model atom, with total
angular momenta $J_{\ell}=0$ and $J_u=1$ for the lower and upper
levels, respectively.
As for the reference values of the $\jkq$ tensors, we consider the
fully converged solution using a very fine GT quadrature with $400$
rays per octant, namely $20$ per inclination and $20$ per azimuth
($20\times 20$).

Our test of the quadratures consists in computing the self-consistent
non-LTE solution using some new quadratures and constructing the
histogram of errors with respect to the reference solution. To perform
this test we use the values of the relevant quantities (i.e. $J^K_Q$)
at the atmospheric height where the optical depth at line centre is
unity for the line of sight along the vertical axis of the model.

Firstly, we test the $S_1$ quadratures presented in Paper~I. We use
the quadrature sets $L=9$, $13$ and $15$ with $8$, $14,$ and $18$ rays
per octant, respectively. We compare such solutions with the
$3\times3$, $4\times4,$ and $5\times5$ GT quadratures. Such sets solve
the system of Eqs. (\ref{eq:system}) up to the same order $L$ as the
$S_1$ quadratures (see Fig.~\ref{fig:nrays}). In
Fig.~\ref{fig:histograms_s1} we show the relative errors of the
radiation field tensors, $\jkq$, for the different quadratures. We see
that both types of quadratures show the same accuracy, while the $S_1$
quadratures contain fewer rays. At $L=9$, both quadratures have a
significant relative error for the tensor components $K\ne 0$,
however, the mean intensity ($J^0_0$) is quite accurately
calculated. The big relative errors in the tensors with $K>0$ indicate
how sensitive the polarisation calculations are to the accuracy of the
angular quadrature.

Once we have checked that the $S_1$ quadratures work well, we can
compare them with the other near optimal quadratures $S_0$ and
$S_2$. We considered the same orders as before, $L=9$, $13$, and
$15$. In Fig.~\ref{fig:histograms_s0_s1_s2}, we show the relative
differences for each tensor component. We see that the quadratures
with the same order provide similar accuracy. Therefore, those with
fewer number of rays are superior even in the 3D calculations. For the
$L=9$ case (see the left panels of
Fig.~\ref{fig:histograms_s0_s1_s2}), we see that the $S_2$ quadrature
provides a somewhat better accuracy than the others, while the
difference disappears in the higher orders. We speculate that this may
be an accidental relic of the particular 3D model we used. This model
is not large enough to be representative of an average atmosphere in a
statistical sense. This can be seen in the difference between the
histograms of Re$[J^2_2]$ and Im$[J^2_2], $ which should look
practically identical provided the model is sufficiently
representative of the average atmosphere.


\section{Discussion and conclusions\label{sec:concl}}

In this paper, we have derived new sets of near optimal angular
quadratures for the numerical solution of 3D radiative transfer
problems of polarised radiation. We have generalised the methods
formulated in Paper~I and the new quadratures we have found are
superior to those published previously.

The new quadrature rules derived in this work have different
transformation symmetries than the previous ones. Imposing the new
symmetries, we can simplify considerably the numerical problem because
the system has fewer unknowns and fewer equations to be solved. In
particular, the new groups of symmetry we have considered are defined
in Eqs.~(\ref{eq:s2}) and (\ref{eq:s3}). In addition, we have
investigated angular quadratures without any particular symmetry.

Such new angular quadratures were calculated using our implementation
of the node elimination algorithm developed by \cite{XIAO2010663}. Our
experience shows that this algorithm is a powerful tool since, in most
cases, the solution converges from any reasonable initial
guess. However, since it is a depth-first search algorithm, the time
complexity of each iteration for a new set of rays is
$\mathcal{O}(N_g)$. This implies that finding a solution for a very
high $L$ can take a significant computing time, especially for angular
quadratures without any kind of symmetries. However, we note that our
algorithms were entirely implemented in Python and much faster solvers
could probably be developed in compiled languages.

The only assumption applied in the derivation of our angular
quadratures is that the angular variation of the radiation field has a
certain level of smoothness. Such an assumption comes into play in
Eq.~(\ref{eq:expansion}), where the finite value of $L$ determines the
maximum order of spherical harmonics up to which the radiation field
can be expanded and for which the calculation of the radiation field
tensors will be exact. A similar limitation is inevitable for any
finite quadrature. However, it is possible to adapt our methods to
find less general quadratures that would be more accurate in
particular astrophysical applications.

The $S_0$ quadratures have proven to be the best option for RT with
polarised radiation. Given the required accuracy, they outperform
other quadratures in terms of number of rays (16\,\% less rays than
$S_1$ and 36\,\% less than the GT quadrature for $L=15$). This is true
both in the analytical calculations, as demonstrated in
Fig.~\ref{fig:nrays}, as well as in realistic 3D simulations, as shown
in Fig.~\ref{fig:histograms_s0_s1_s2}. The tables with the different
sets of the $S_0$ quadratures, both for the polarised case and for
just the specific intensity, are available at the CDS.

It is not surprising that quadratures with a very high degree of
symmetry cannot compete with the most general setup. However,
quadrature sets with some kind of symmetry, such as that of the $S_1$
set, can be significantly easier to find numerically even for much
larger values of $L$ than we have considered in this work. However,
robust techniques, such as the node elimination algorithm, need to be
employed instead of a naive brute-force solution.

In our calculations, we forced the rays to avoid any octant edges,
the poles, and the equator. Our argument for this is purely technical
given that in some of the RT codes, such as PORTA, it is convenient to
avoid such types of rays. This restriction does not affect the $S_0$
quadratures because they are not defined with respect to the octant
boundaries, but it may, at least in principle, create unnecessary
constraints for the other sets.

\begin{acknowledgements}
  J.J.B. acknowledges financial support from the Spanish Ministry of
  Economy and Competitiveness (MINECO) under the 2015 Severo Ochoa
  Programme MINECO SEV--2015--0548. J.\v{S}. acknowledges the
  financial support of grant \mbox{19-20632S} of the Czech Grant
  Foundation (GA\v{C}R) and from project \mbox{RVO:67985815} of the
  Astronomical Institute of the Czech Academy of
  Sciences. J.T.B. acknowledges the funding received from the European
  Research Council (ERC) under the European Union's Horizon 2020
  research and innovation programme (ERC Advanced Grant agreement
  \mbox{No.~742265}), as well as through the projects
  \mbox{PGC2018-095832-B-I00} and \mbox{PGC2018-102108-B-I00} of the
  Spanish Ministry of Science, Innovation and
  Universities. J.\v{S}. and J.T.B. acknowledge the support from
    the Swiss National Science Foundation through the Sinergia grant
    \mbox{CRSII5-180238}. The authors thankfully acknowledge the
  technical expertise and assistance provided by the Spanish
  Supercomputing Network (Red Espa\~nola de Supercomputaci\'on), as
  well as the computer resources used: the MareNostrum supercomputer
  in Barcelona and LaPalma Supercomputer.
\end{acknowledgements}

\bibliography{mybibtex}{}
\bibliographystyle{aa}



\end{document}